\documentclass{amsart}
\usepackage{amsaddr}
\usepackage{natbib} 
\usepackage{graphicx}
 \usepackage{caption}
 \usepackage{subcaption}
\usepackage{geometry}                
\usepackage{graphicx}
\usepackage{amssymb}
\usepackage{epstopdf}
\usepackage{cancel}

\title{A general framework for modelling zero inflation}
\author[Haslett et al, A general framework for modelling zero inflation]{John Haslett$^1$, Andrew Parnell$^2$, James Sweeney$^3$}
\address[A1]{$^1$School of Computer Science, Trinity College Dublin}
\address[A2]{$^2$Hamilton Institute, Maynooth University}
\address[A3]{$^3$School of Business, University College Dublin}

\newcommand {\nc} {\newcommand}

\nc {\Stthfn}[1] {\tilde S_{\theta}\big( #1\big)}
\nc {\Stgmfn}[1] {\tilde S_{\gamma}\big( #1\big)}

\nc {\Stththfn}[1] {\tilde S_{\theta\theta}\left( #1\right)}
\nc {\Sthfn}[1] {S_{\theta}\left(#1 \right)}
\nc {\Sththfn}[1] {S_{\theta\theta}\left(#1 \right)}
\nc {\Stomfn}[1] {\tilde S_{\omega}\left(#1 \right)}
\nc {\Stomomfn}[1] {\tilde S_{\omega\omega}\left(#1 \right)}
\nc {\Stthomfn}[1] {\tilde S_{\theta\omega}\left(#1 \right)}
\nc {\Stbetafn}[1] {\tilde S_{\beta}\left(#1 \right)}
\nc {\Stalphafn}[1] {\tilde S_{\alpha}\left(#1 \right)}

\nc \yb {\mathbf{y}}
\nc \Yb {\mathbf{Y}}
\nc \xb {\mathbf{x}}
\nc \Xb {\mathbf{X}}
\nc \fb {\mathbf{f}}
\nc \Mb {\mathbf{M}}

\nc \EE {\mathrm{I\!E\!}}
\nc \RR {\mathrm{I\!R\!}}
\nc\ZZ {\mathbb{Z}}

\nc \yd {\dot{ y} }
\nc \ydd {\ddot{y}}
\nc {\Yt} {\tilde Y}
\nc {\Ybt}{\tilde {\mathbf Y}}
\nc {\Jt} {\tilde J}
\nc {\kp}{\kappa}
\nc {\pit} {\tilde \pi}
\nc{\taut}{\tilde \tau}
\nc {\Iy} {\mathbb{I}\{y=0\}}
\nc {\IY} {\mathbb{I}\{Y=0\}}
\nc {\IYt}{\mathbb{I}\{\Yt=0\}}
\nc {\IYbt}{\mathbb{I}\{\Ybt=0\}}
\nc {\IYti}{\mathbb{I}\{\Yt_i=0\}}
\nc {\IYtj}{\mathbb{I}\{\Yt_j=0\}}
\nc {\IYj}{\mathbb{I}\{Y_j=0\}}
\nc {\IYtN}{\mathbb{I}\{\Yt=N\}}
\nc {\IYto}{\mathbb{I}\{0<\Yt<N\}}
\nc {\Iyi} {\mathbb{I}\{y_i=0\}}
\nc {\Iyni} {\mathbb{I}\{y_i \neq 0\}}
\nc {\Iyih} {\mathbb{I}\{y_i=h\}}
\nc{\nz}{\cancel{0}}

\nc{\logit}{\mbox{logit}}
\nc{\expit}{\mbox{expit}}

\nc {\StthYt} {\tilde S_{\theta} (\theta, \gamma; \tilde Y)}
\nc {\SthY}{S_{\theta} (\theta; Y)}
\nc {\Stthy} {\tilde S_{\theta}  \theta, \gamma ; y}
\nc {\Stthyi} {\tilde S_{\theta} ( \theta_i, \gamma_i ; y_i)}

\nc {\Stalphay} {\tilde S_{\alpha} (\theta, \kappa; y)}
\nc {\Stom} {\tilde S_{\omega}}
\nc {\Stq} {\tilde S_{q}}
\nc {\StkpYt} {\tilde S_{\kappa}(\omega \theta ; \tilde Y)}
\nc {\StbetaYt} {\tilde S_{\beta} (\omega, \beta \omega; \tilde Y)}
\nc {\Sbetaj} {S_{\beta_j}}
\nc {\Stbeta} {\tilde S_{\beta}}
\nc {\Stalpha} {\tilde S_{\alpha}}
\nc {\Stgm} {\tilde S_{\gamma}}
\nc {\Stbetaj} {\tilde S_{\beta_j}}
\nc {\Stalphahj} {\tilde S_{\alpha_h}}
\nc {\StbetajYt} {\tilde S_{\beta_j} (\beta, \omega; \tilde Y)}
\nc {\Stbetajy} {\tilde S_{\beta_j} (\omega; y)}
\nc {\StomYt} {\tilde S_{\omega} (\theta, \omega; \tilde Y)}
\nc {\StomhYt} {\tilde S_{\omega_h} (\theta, \omega; \tilde Y)}
\nc {\Stomy} {\tilde S_{\omega} (\theta, \omega; y)}
\nc {\Stth}{\tilde S_{\theta}}
\nc {\Sthy}{S_{\theta}(\theta; y)}
\nc {\Sthyi}{S_{\theta_i}(\theta_i; y_i)}
\nc {\Sthz}{S_{\theta}(\theta; 0)}
\nc {\Sthzi}{S_{\theta_i}(\theta_i; 0)}

\nc{\ddthfn}[1]{\frac{\partial #1}{\partial \theta}}
\nc{\ddkpfn}[1]{\frac{\partial #1}{\partial \kappa}}
\nc{\ddomfn}[1]{\frac{\partial #1}{\partial \omega}}
\nc{\ddgmfn}[1]{\frac{\partial #1}{\partial \gamma}}
\nc {\ddbetajfn}[1] {\frac{\partial #1 }{\partial \beta_j}}
\nc{\ddafn}[1]{\frac{\partial #1}{\partial {\alpha}}}
\nc{\ddahfn}[1]{\frac{\partial #1}{\partial {\alpha_h}}}
\nc{\ddqfn}[1]{\frac{\partial #1}{\partial {q}}}
\nc {\ddkp} {\frac{\partial  }{\partial \kappa}}
\nc {\ddth} {\frac{\partial  }{\partial \theta}}
\nc {\ddthi} {\frac{\partial  }{\partial \theta_i}}
\nc {\ddpiz} {\frac{\partial  }{\partial \pi_0}}
\nc {\ddkpi} {\frac{\partial  }{\partial \kappa_i}}
\nc {\ddkh} {\frac{\partial  }{\partial \kappa_h}}
\nc {\ddom} {\frac{\partial  }{\partial \omega}}
\nc {\ddomh} {\frac{\partial  }{\partial \omega_h}}
\nc {\ddpifn}[1] {\frac{\partial  #1}{\partial {\pi_0}}}
\nc {\ddbetaj} {\frac{\partial  }{\partial \beta_j}}
\nc {\ddbetak} {\frac{\partial  }{\partial \beta_k}}
\nc {\ddbeta} {\frac{\partial  }{\partial \beta}}
\nc {\ddalpha} {\frac{\partial  }{\partial \alpha}}
\nc {\dthbetaj} {\frac{\partial  \theta}{\partial \beta_j}}
\nc {\ddomsq} {\frac{\partial^2  }{\partial \omega^2}}
\nc {\ddthsq} {\frac{\partial^2  }{\partial \theta^2}}
\nc {\ddthom} {\frac{\partial^2  }{\partial \theta \partial \omega}}
\nc {\ddombetaj} {\frac{\partial^2  }{\partial \beta_j \partial \omega}}
\nc {\ddbetajk} {\frac{\partial^2  }{\partial \beta_j \partial \beta_k}}

\nc{\grad} {\nabla}

\nc{\FItfn}[1]{\tilde{\mathcal{I}}_{#1}}
\nc{\FIt}{\tilde{\mathcal{I}}_{ \theta, \gamma}}
\nc{\FIti}{\tilde{\mathcal{I}}_{ \omega, \theta_i}}
\nc{\FI}{\mathcal{I}_{\theta}}
\nc{\FIbeta}{\mathcal{I}_{\beta}}
\nc{\FItbeta}{\tilde{\mathcal{I}}_{\beta, \omega}}
\nc{\FItbetom}{\tilde{\mathcal{I}}_{\beta, \omega}}
\nc{\FIomega}{\mathcal{I}_{\omega}}

\begin{document}
\maketitle

\section*{Abstract}

We propose a new framework for the modelling of count data exhibiting zero inflation (ZI). The main part of this framework includes a new and more general parameterisation for ZI models which naturally includes both over- and under-inflation. It further sheds new theoretical light on modelling and inference and permits a simpler alternative, which we term as multiplicative, in contrast to the dominant mixture and hurdle models. Our approach gives the statistician access to new types of ZI of which mixture and hurdle are special cases. We outline a simple parameterised modelling approach which can help to infer both ZI type and degree and provide an underlying treatment that shows that current ZI models are themselves typically within the exponential family, thus permitting much simpler theory, computation and classical inference. We outline some possibilities for a natural Bayesian framework for inference; and a rich basis for work on correlated ZI counts.\\
 
The present paper is an incomplete report on the underlying theory. A later version will include computational issues and provide further examples.\\

\section{Introduction}

We consider regression modelling of observed count data $\yb= \{y_1, \ldots, y_n\}$ on covariates $\Xb = \{\xb_1, \ldots, \xb_n\}$ where many values of $y_i$ are 0; termed zero-inflation (ZI). We treat the count data as realisations of a random variable $\Yt$, having pmf $\pit_y = \pit_y(\theta, \kappa), y \in \ZZ^+_0$, with the parameters $\theta$ and $\kappa$ controlling the location and ZI processes respectively, with e.g. $\theta_i = \theta(\xb_i\beta)$. The pmf $\pit_y$ is related to a simpler pmf $\pi_y = \pi_y(\theta)$ which characterises the standard count process, absent ZI; typically $\pit_0 > \pi_0$. Observations on $\Yt$ thus imperfectly reflect $\pi_y(\theta)$. We propose a new approach for the two main issues associated with such models: (a) performing inference on $\theta$ and $\kappa$ given a specific $\pit_y$ and (b) constructing families of $\pit_y$ from families of $\pi_y$.  Our main contribution is the modelling of the function $\pit_0 = \pit_0\big(\pi_0(\theta), \kappa\big)$ which allows us to diagnose a wide class of zero inflation types, some of which have not been identified in the literature to-date. Figure  presents examples of such types. However we also identify a universal system of equations for (a).
\\

\begin{figure}[h!]
  \includegraphics[width=\linewidth]{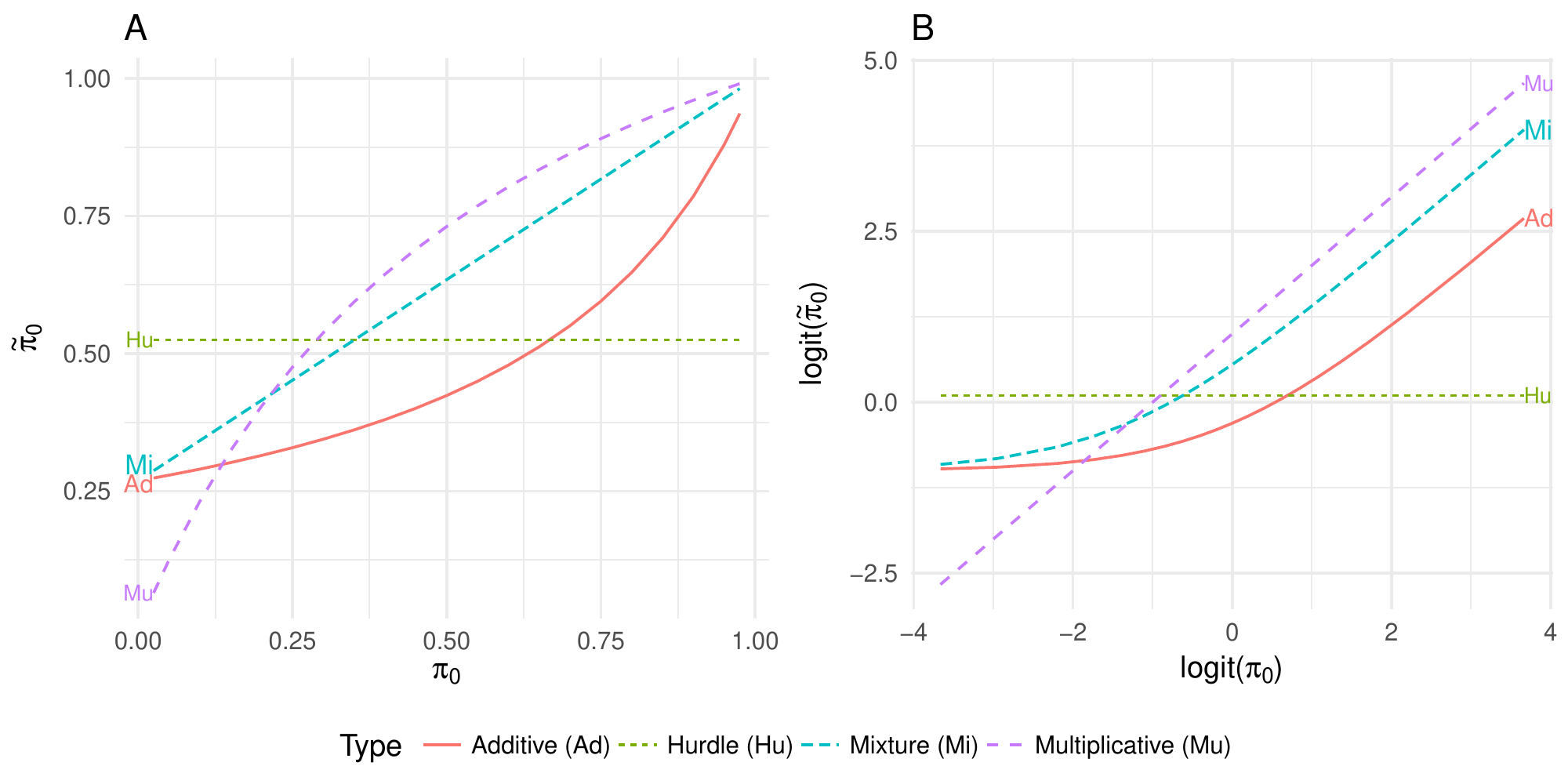}
  \caption{Multiplicative, mixture, hurdle and additive models. A: probability metric. B: logit metric}
  \label{fig:coffee}
\end{figure}

The seminal types of ZI are those of the so-called hurdle model \citep{Mullahy1986hurd1, Mullahy1997hurd2}, and the mixture model of \citet{Lambert1992:ZIP}. 
In the hurdle model, all instances of $\Yt = 0$ are generated by a Bernoulli process with $\pit_0$, such that $\pit_0$ is not a function of $\pi_0$; hence observed 
zeroes contain no information on $\theta$. Conversely, all instances of $\Yt > 0$ are generated by a distribution $\pi_y$, defined solely on $y>0$; typically a 
truncated distribution. In the mixture model, there are two latent variables: $Y$ as defined above together with binary $J$ such that $P(J=1) = q$, where 
$q = q(\kappa)$. The observable is $\Yt=YJ$. Now $\pit_0 = (1-q) + q \pi_0$ hence $\pit_0$ is a linear function of $\pi_0$ and a function of $\theta$. 
In the mixture model only some instances of  $\Yt=0$ are relevant to inference on $\theta$; the challenge is that we do not know which.\\

Since their publication, the literature has generated more than 1000 papers\footnote{Google Scholar search for ``zero inflation'' retrieved on 25/4/18} with very 
many applications of these two models; excess zeroes arise in many contexts. There have been technical extensions, such as algorithms for mixed models for 
which the seminal paper is  \citet{Hall2000ZI}; the use of distributions other than the simple Poisson and binomial used in the early papers, and in particular of 
distributions that facilitate the separation of over-dispersion and zero-inflation, including \citet{ridout2001NegBin}, \citet{xiang2007score}, and  
\citet{kassahun2014ZI:overdisp}. Several authors have pursued Hall's lead on random effects, including (as well as some of the above) \citet{long2015random}, 
\citet{min2005random}, \citet{martin2017marginal}, and \citet{molas2010hurd-mix}. These of course have much in common with other examples of correlated, zero-
inflated data, such as arise in studies with a focus on longitudinal, time-series and spatial data, including \citet{chebon2017:too.complex}, \citet{yang2016long}, 
\citet{agarwal2002ZIspatial}, \citet{ancelet2010ZIspatial}. Multivariate data exhibiting ZI have been examined in papers such as \citet{li1999multivarZI} and 
\citet{liu2015typeI}. Several authors have pursued Bayesian inference in ZI, including \citet{angers2003bayesian}, \citet{rodrigues2003bayesian}, 
\citet{dagne2004Bayes}, \citet{ghosh2006bayesian}, \citet{klein2015}, \citet{lee2017}, \citet{neelon2017:LZIP} and some of those cited earlier. Thus is a very active area of research.\\

There are several useful reviews. \citet{ridout1998models} point out, inter-alia, that: (i) the hurdle and mixture models can in fact be seen as re-parameterisations of each other; and (ii) that the parameterisation $\pit_0 = (1-q) + q \pi_0$, although suggested by the mixture model, in fact allows $\kappa>1$ (subject to $\pit_0 \in [0,1]$); this can characterise zero deflation. \citet{ghosh2012:kZIG} draw attention to early precursors, such as  \citet[][although there does seems to be a citation error]{cohen1960}. Both \citet{ridout1998models} and \citet{ghosh2012:kZIG} emphasise that these ideas suggest different parameterisations; for in the simplest \emph{iid} case all ZI models are equivalent by re-parameterisation. But, remarkably, there seem to be few if any attempts to set models of zero inflation in a wider modelling framework. In this paper we distinguish between (i) regression modelling of counts on covariates in the presence of ZI; and (ii) modelling of zero-inflation, per se. The latter  focusses on both the type and degree of ZI, possibly also involving covariates. This modelling of ZI seems, surprisingly, to be completely unexplored.\\

The hurdle and mixture models are well defined in an abstract sense, or for Monte Carlo simulations. But are they really rich enough to complement  and stimulate  the more process based thinking of many subject matter specialists? Might they feel, at  least sometimes, that they are being shoe-horned by the statistical community into an unnatural framework?   \citet{chebon2017:too.complex} raise the query in their title: ``Models for zero-inflated, correlated count data with extra heterogeneity: when is it too complex?''  \citet{todem2016} remark ``much of the literature on real applications of these models has centered around the latent class formulation where the mean response of the so-called at-risk or susceptible population and the susceptibility probability are both related to covariates. While this formulation in some instances provides an interesting representation of the data, it often fails to produce easily interpretable covariate effects on the overall mean response."\\

Several applied authors register similar anxieties. \citet{miller2008} report ``The results suggest that the best-fitting zero-inflated model sometimes depends on the proportion of zeros and the distribution for the non-zeros. In fact, there are situations where the zero- inflated models are not necessary. \citet{garay2011diagnostics} says: ``In order to study departures from the error assumption as well as the presence of outliers, we perform residual analysis .....illustrated with a real data set, where it is shown that, by removing the most influential observations, the decision about which model is best as the data changes. Similar is \citet{fisher2017managingZI}: ``Although these models are often appropriate on statistical grounds, their interpretation may prove substantively difficult."  This all suggests an un-met need for a wider range of ZI models easy to interpret and to criticise. \\

The first thought in such modelling is surely the question of whether, and if so \emph{why},  there is a need to model zeroes differently. Is this need always well served by the two main models? We will argue that they are not even the simplest. At a more  technical level, we observe that in the Mixture model, the null model corresponds to $q=1$,  an extremum in the parameter space. Are there natural model variations which are well defined in a region \emph{around} the null?  \\

The hurdle model is often described as permitting both zero inflation and zero deflation. But it is not strictly necessary that the underlying distribution $\pi_y$, truncated to have support only on $y>0$, have any interpretation at $y=0$. The common use of truncated distributions is most often a convenient way to marry extant statistical machinery to the necessity of a model defined only on $y>0$. Its use does not always derive from a need to question, in a natural way, the evidence that zeroes are different. The null model is not - in any clear sense - nested within the hurdle model unless a truncated $\pi_y$ is used; then, for $y>0, \pit_y =\frac{1-\pit_0}{1-\pi_0}\pi_y$; but $\pit_0$ does not depend on $\theta$. In this paper, we take the hurdle model to be defined with respect to a truncated distribution.\\

\citeauthor{Lambert1992:ZIP}'s \citeyear{Lambert1992:ZIP} formulation led to an EM algorithm, which dominates the implementation of the mixture model: the variable $J$ is taken as providing the complete data likelihood. The algorithm leads to (a) down-weighting zeroes to estimate regression coefficients; and (b) a clever, but highly technical, use of binary regression of latent $J$  on covariates, in order to estimate  $\kappa$. We will see (a) does not need to appeal to EM for motivation, as it flows directly from maximum likelihood; on (b) we propose alternatives which do allow natural criticism of this choice of ZI model.\\

Lambert enters a caveat; she remarks that the the EM algorithm cannot be used if $\theta$ and $\kappa$ are related, including via a dependence on common covariates. She is clear that the caveat is directed at the EM algorithm, rather than at the use of maximum likelihood \emph{per se} to estimate the parameters of the Mixture model. She also remarks that in the event that (modulo notation) ``If the same covariates affect $\kappa$ and $\theta$, it is natural to reduce the number of parameters by thinking of $q$ as a function of $\kappa$''.  This saves computation, as she says. But it is of course perfectly sensible advice for seeking to simplify the statistical model, the rewards for which go well beyond computation. One parsimonious example is \citet{salter2012}, where
$q$ is modelled as  $\left(\frac{\mu_y(\theta)}{1+ \mu_y(\theta)}\right)^{\gamma}$, where $\gamma$ is a scalar characterising the degree of ZI, and 
$\mu_y(\theta)$ is the expected value of the pmf $\pi_y$.
One conclusion of the current paper is that the mixture model is simple only in a rather specific sense.\\

It has been noted that ZI can be difficult to distinguish from over-dispersion (\citet{perumean2013zero}). This has led several researchers to  build ZI models $\pit_y$ on richer examples of $\pi_y$. Researchers have considered the Negative Binomial, which can be considered as mixture of Poissons; e.g. \citet{ridout2001NegBin},  \citet{moghimbeigi2011score}, and \citet{garay2011diagnostics}. Others have used the Conway-Maxwell-Poisson distribution \citep{sellers2016disp} and Generalised Poisson distributions \citep{xie2009score}. But these are all specific to one example of $\pi_y$, the Poisson. Several ZI researchers have found it fruitful to use distributions from the power series family \citep[e.g.][]{bhattacharya2008,patil2011tests}. Bizarrely there seems to be little interest in building $\pit_y$ from more classic families of $\pi_y$, such as the classic exponential-dispersion family of \citet{jorgensen1987expdisp}, or the over-dispersed exponential family \citep{gelfand1990,dey1997overdisp:glm}. We shall see below that our simplest ZI model fits very naturally with the exponential family.\\

In this paper we provide an apparently new and much wider framework for ZI modelling, which we distinguish from regression in the presence of ZI. Here, because its widespread acceptance, we use the ZI label to include various versions of zero-modification \citep[e.g.][]{min2010ZeroMod}, including both over- and under-inflation (deflation). We focus on the univariate case, and work primarily within the exponential family, extending to multivariate in later sections. We concentrate on inference through the likelihood, primarily because it provides interesting insights on the options for ZI modelling, but the general ideas extend naturally to other frameworks. The key contribution lies in the next two sections. We hope that the framework will open new avenues for others to pursue. \\

\section{A general model for ZI}

We adopt the notation
\begin{equation*}\label{basic}
\pit_y(\kappa, \theta) = \left\{ \begin{array}{ll} (1+\kappa)\rho \pi_0(\theta) & \mbox{if } y = 0 \\ \rho \pi_y(\theta) & \mbox{otherwise}. \end{array} \right.
\end{equation*}
where the function $\rho = \rho( \pi_0,\kappa)$ renormalises and $\kappa$ controls ZI as before. Typically $\pit_0 > \pi_0$ corresponding to $\kappa >0$ and $0<\rho <1$. However also permissible is $-1 < \kappa < 0$ which implies $\pit_0 < \pi_0$ and $\rho > 1$. The simplest expression for $\rho$ is $\rho = \frac{1-\pit_0}{1-\pi_0} $, from which it follows  that 
\begin{equation*} \label{OR}
\frac{\pit_0}{1-\pit_0} = (1+\kappa)\frac{\pi_0}{1-\pi_0} \hspace{1 cm}
\mbox{or equivalently, }\hspace{1cm}
\logit(\pit_0) = \omega + \logit(\pi_0)
\end{equation*}
with $\omega = \log(1+\kappa)$. We refer to these as the odds ratio and the log-odds forms of ZI respectively. It is a simple matter to show that $\rho^{-1} = (1 + \kappa \pi_0) = (1-\pi_0) + e^{\omega} \pi_0$. Note that central to the notation is an explicit functional relationship between $\pit_0$ and $\pi_0$, characterised by $\omega$. Below we shall treat $\omega$ as a function of $\pi_0$, that is $\omega = \omega(\gamma, \pi_0)$ where $\gamma$ is a scalar parameter controlling the degree of ZI. In regression (where $\theta$, and thus $\pi_0(\theta)$,  and also $\gamma$ may vary with covariates) the choice of function is at the heart of the modelling of the type of ZI. A wide class of ZI models can be obtained this way.\\

The relationship between $\pit_y$ and $\pi_y$ may be more compactly written as
\begin{equation*}
\pit_y = (1+\kappa )^{\Iy} \rho \pi_y; \mbox{ or equivalently }
\log (\pit_y) = \omega {\Iy} + \log (\rho) + \log (\pi_y) 
\end{equation*}
Note the duality between $\omega$ and $\rho$. Each defines the other via normalisation; but a simple parametrisation for one may require a difficult parameterisation for the other. The simplest model in this notation is that in which $\omega$ is constant wrt $\pi_0$; but $\rho$ is thus dependent on $\pi_0$. In contrast, as shown later, the mixture model has $\rho$ independent of  $\pi_0$; this leads to $\omega$ being dependent on $\pi_0$.\\

A natural and generic simulation mechanism for $\Yt$ is as below, with $U$ denoting a realisation from $U(0,1)$:
\begin{enumerate}
\item If $U \le \pit_0$ generate $\Yt = 0$
\item  Otherwise generate $\Yt$ from $\pi_y$, rejecting all instances of 0
\end{enumerate}
Thus in the latter case, with probability $1- \pit_0$ we sample from the truncated pmf $\frac 1 {1- \pi_0} \pi_y$. Observe that there are no latent variables involved in the data generation. This constructive formulation makes it apparent that the central parameters, for simulation and inference, are $(\pit_0, \theta)$. \\

The formulation we adopt includes, as a special case, the hurdle model, where $\omega =\gamma-\logit(\pi_0)$, and thus $ \logit(\pit_0) = \gamma$ is independent of $\pi_0$. A further special case is the classic mixture model; here $\pit_0 = (1-q) + q\pi_0$, and $\pit_y = q\pi_y, \forall y>0$ with $q$ the mixture weight. We may identify $q$ with $\rho$, noting in particular that, in this ZI model, the normalising function $\rho(\cdot)$ is typically modelled as independent of $\pi_0(\theta)$  \emph{per} Lambert's caution on EM. From this identification, with $e^{-\gamma} = \frac{1-q}{q}$ we find $\kappa = \frac{e^{-\gamma}}{ \pi_0}$; hence $\omega = \log(\pi_0+e^{-\gamma}) - \log(\pi_0)$  is a function of $\pi_0(\theta)$; and in this sense it is not as simple as constant $\omega$. However, $\pit_0$ is a linear function of $\pi_0$. These, and other, types of ZI are discussed below.\\
 
 
Further, the latent variable definition of the mixture model admits the very simple interpretation of positive $\kappa$; for $P(J = 1 | \Yt=0) =\frac{q\pi_0}{\pit_0}= \frac 1 {1+\kappa} = e^{-\omega}$. This probability is central to Lambert's EM algorithm, as we elaborate below. But we will see below that EM is natural only in the very specific context of the mixture model of ZI. Of some interest below will be the interpretation, as an expected value, rather than a probability, of $\frac 1 {1+\kappa} = e^{-\omega}$. It is thus simply interpretable for negative $\omega$ also, corresponding to $-1 < \kappa < 0$.\\

Under-inflation of zeroes can be thought of via probabilistic censoring. Consider the following data generating model: (i) generate $Y$ from $\pi_y(\theta)$; (ii) return $\Yt=y$ if $y >0$; but, (iii) when $y=0$, return a value of $missing$,  with probability $r$. The pmf of the \emph{non-missing} $\Yt$  is $\pit_y$, and $P(\Yt=0 | not\,missing) = \pit_0 = \frac{\pi_0(1-r)}{1-\pi_0r}$. Thus, with $\kappa \in (-1,0)$, we may identify $\kappa$ with $-r$; note that the limiting case of $\kappa \to -1$ (or equivalently, $\omega \to -\infty$) corresponds to  truncation at  $y=0$. \\

Equivalently, we may associate with each observed $\Yt=0$ an unobserved random (integer) number $M$ of instances of $Y=0$, of which the observed zero is the sole survivor. Then (with $\kappa<0$) $P(M=m) = (-\kappa)(1+\kappa)^{m-1}; m \in \ZZ_0^+$ with
$E[M] = \frac 1 {1+\kappa}$; but $M$ is only defined when $\Yt=0$. We note that the same interpretation can apply to over-inflation, but now $M$ is binary; $M=0$ here corresponds to $J=0$, an `inflated' zero. Now $E[M] = \frac 1 {1+\kappa}$, a value shared with $E[J|\Yt=0]$ in the mixture model. Note that in both cases $Var[M] = \frac{\kappa} {1+\kappa}$. Thus in one sense the variable $M$ is a generalisation of the variable $J$ which defines the mixture model. \\

Despite $M$ being undefined if $\Yt \ne 0$, it is natural to extend its definition, for we can write $M=1$ when $\Yt \ne 0$. Then $E[M | \Yt] = \left(\frac 1 {1+\kappa}\right)^{\IYt}$ covering all cases. The unconditional distribution of $M$ is $\pi_M(m) = \sum_y P(M=m | \Yt=y)\pit_y; m \in \ZZ_0^+$. We consider separately the cases of positive and negative $\kappa$. With $\kappa >0$, $\pi_M(m)= 0$ for $m>1$. Then $\pi_M(0) = \frac{\kappa}{1+\kappa}\pit_0= \frac{\kappa}{1+\kappa}(1+\kappa)\rho \pi_0 = 1-\rho$, and $\pi_M(1)= \rho$; when $\rho=q$, as in the mixture model, this coincides with the distribution of $J$. For negative $\kappa$, we note that $m>1$, $P(M=m | \Yt=y) = 0$ unless $y=0$; thus $\pi_M(m) = \pit_0(-\kappa)(1+\kappa)^{m-1} = (\rho -1)(1+\kappa)^m$, when $m>1$. It follows that $\pi_M(1)  = -\kappa^{-1}(\rho-1)(1+\kappa)^2$. \\

The latent variable $M$ is best understood as the number of instances  of (unobserved) $Y=0$ associated with every instance of observed $\Yt=0$. It will be noted therefore that, given $n_0$ \emph{iid} instances of  $\Yt=0$, then, on average, $\frac {n_0} {1+\kappa}$  are relevant to inference on $\theta$. In a sample of $n$  \emph{iid} observations, the expected number of zeroes is $n\pit_0$, so the expected total number in the sample, relevant to such inference, is  $n\left( 1 - \frac {\kappa \pit_0} {1+\kappa}\right) = n\rho$. This may be thought of as an effective sample size, being greater or less than $n$ for under-and over-inflation, respectively; this echoes \citet{cohen1960}.  We formalise this below in the context of $\pi_y(\theta)$ in the exponential family. But note that $M$ is not part of the definition of our ZI model, in contrast to the role of $J$ in the classic mixture model.\\
 
\subsection{Types of ZI}\label{ZItypes}

From the parameterisation above, we see that $\pit_0$, as a function of $\pi_0$, is most simply expressed via the log-odds form $\logit(\pit_0)= \omega + \logit(\pi_0)$. The implicit function $\omega\Big(\gamma(\xb \alpha), \pi_0(\theta(\xb \beta))\Big)$ characterises the type of ZI, with any parameters $\alpha$ defining the degree. The functions $\theta(\cdot)$ and $\gamma(\cdot)$ are functions relating the covariates to $\tilde \mu = E[\Yt]$; $\gamma(\cdot)$ (and hence $\alpha$) characterises degree of the ZI, within the type specified by the function $\omega(\gamma, \pi_0)$. In our simplest models we will typically consider these to be identity functions. Indeed the function $\omega(\cdot)$ only becomes visible in the presence of covariates $x$. For then, within a data set, observations $y_i$ are associated with different $\theta_i = \theta(\xb_i\beta)$ and $\gamma_i = \gamma(\xb_i\beta)$, and thus with pairs $(\pit_{i0}, \pi_{i0})$, these being defined as $\pi_{i0} = \pi_0(\theta_i)$ and   $\pit_{i0} = \pi_0(\theta_i, \omega(\gamma_i, \pi_0))$. Types of ZI are characterised by  functions such as $\pit_0(\gamma, \pi_0)$ that are defined by a given function $\omega$; the scalar parameter $\gamma$ controls the degree of ZI. \\

The simplest model has $\logit(\pit_0)-\logit(\pi_0) =\log(1+ \kappa) = \omega =  \gamma$; here the function $\omega$ is constant with respect to $\pi_0$. It appears to be new. We refer to it as multiplicative ZI. It is strictly multiplicative only in the odds ratio sense; that is, ratios $\frac{\pi_{0}}{1-\pi_{0}}$ are increased (or decreased) to $\frac{\pit_{0}}{1-\pit_{0}}$ by a multiplicative factor. But we note that when $\pi_0$ is small (corresponding to parts of the covariate space where 
$\pi_0(\theta(\xb \beta))$ is small) so necessarily must be $\pit_0$; we may argue similarly for large $\pi_0$ and  $\pit_0$. In these circumstances, the nature of ZI is to accentuate the variation in $\pi_{i0}$ that is induced by varying $x_i$ in covariate space; it is in the sense of `accentuate' that we use the term `multiply'. 
In the context of Figure 1, when $\pi_{i0}$ is very small, or very large, so also will be  $\pit_{i0}$; but when $\pi_{i0} \approx 0.5$, $\pit_{i0}$ will be larger than $0.5$, for positive $\omega$, and smaller than $0.5$ for negative $\omega$.
It is also uniformly inflationary in the sense that, for $\omega > 0$, $\pit_0(\omega, \theta) > \pi_y(\theta)$ for all $\theta$, and conversely for $\omega < 0$.\\

As seen above, the classic mixture model above is another type  of ZI, characterised here by the rather more awkward $\omega = \log(\pi_0+e^{-\gamma}) - \log(\pi_0)$, where the awkward additive terms in the first expression reflect the typical difficulties of a mixture model. It is these difficulties that lead to some awkward computations, for which  EM  here can  supply solutions. We may describe this model as having additive ZI, in the sense that even in parts of covariate space where $\pi_0$ is small, $\pit_0$ can be such that many zeroes are observed. As remarked, with $q >1$ (subject to $0 \le \pit_y \le 1$), this type of ZI can be under-inflationary. This constraint is difficult to parameterise, however, and $q>1$ is rarely used. This model is in fact also uniformly inflationary in the sense of the previous paragrpah.\\

Another well studied type is the hurdle model. This need not be cast as either over- or under-inflation, for $\pi_y(\theta)$ is only defined on $y > 0$. However, here we  only use truncated versions of distributions such as the Poisson. In these circumstances we can say that  $\pi_0$ is defined, at least implicitly, and the distinguishing  feature of the truncated model is thus that, although $\pit_0$ is well defined, it is constant \emph{wrt} $\pi_0(\theta)$. In our notation, the (truncated) hurdle type of ZI corresponds to  $\omega =e^{\gamma} - \logit(\pi_0)$. This ZI type is, unsurprisingly, not uniformly inflationary.\\

An apparently new type is available through $\omega = \xb \alpha - \log(\pi_0)$. Now $\logit\left(\pit_0\right) =  \xb \alpha -\log(1-\pi_0)$, an increasing function of $\pi_0$. This has the property that, for small $\pi_0$, $\pit_0 \approx \frac{e^{\xb \alpha}}{1+e^{\xb \alpha}}$. In this sense this parameterisation also has the additive property. We refer to it as additive, recognising however that that mixture model also has this property. As seen in Figure 1, it is not uniformly inflationary.\\

More generally, the multiplicative, additive, and hurdle are all special cases of $\omega = \gamma + \tau_1 \log(\pi_0) + \tau_2 \log(1-\pi_0)$; they correspond to $(\tau_1, \tau_2) = (0,0), (0,-1), (-1,-1)$, respectively. In this paper, apart from the classic mixture model, we shall restrict ourselves to ZI types of this form. However, the reader will note that there is no general restriction in ZI type; the function $\omega(\gamma, \pi_0)$ may be modelled very flexibly. We may, for example, extend our notation $\gamma$ to refer both to coefficients of $\xb$ in the link function $\gamma(\alpha \xb)$ and to coefficients of the chosen functions of $\pi_0$, such as the log functions used above. We shall take the $\tau$ coefficients as known,  unless otherwise stated; but they too can be estimated from data. \\

Other functions $\pit_0(\gamma, \pi_0)$, equivalently other ZI types, can of course be defined as convenient. But their key properties are characterised by the function $\omega = \logit(\pit_0) - \logit(\pi_0)$. For example, the recursive use of ZI by \citet{ghosh2012:kZIG} corresponds to a new distribution $\tilde {\pit}_y$ being related to $\pit_y$ via the odds ratios:
\begin{equation*}
OR(\tilde {\pit}_0)= (1+ \tilde \kappa)OR(\pit_0)=  (1+ \tilde \kappa) (1+  \kappa)OR(\pi_0) =  (1+ \tilde {\tilde \kappa})OR(\pi_0)
\end{equation*}
or equivalently via $\tilde {\tilde \omega} = \tilde \omega + \omega$, concatenating the two inflations. As the authors remark, these functions cannot be identified separately; but using different parameterisations for each is one way  to motivate a new parameterisation for a single function $\omega$.\\

Critically the type of ZI depends on the functional relationship between $\pit_0$ and $\pi_0$, moderated by the covariates $\xb$ and captured by $\omega(\cdot)$. This is where lies the essence of ZI modelling, which we may contrast with  regression of $y$ on $x$ in the presence of ZI. This relationship can be arbitrarily rich and need not be linear; but simplicity usually brings more insight. \\

We shall write 
$$\omega = \logit(\pit_0)-\logit(\pi_0) = \gamma(\alpha \xb) + \sum_k \tau_k f_k(\pi_0)$$
where the functions $f_k(\pi_0)$ typically include the log functions $\log(\pi_0), \log(1-\pi_0)$ and the $\tau_k$ are coefficients independent of  $\xb$. When these are known, the ZI model is specified. We shall see that linear logistic regression of $\Iy$ on $\xb$ and (if the ZI type is not known) on a small number of functions $f_k(\pi_0)$ suffice to estimate the coefficients $\alpha$ and $\tau$. Compactly we can write this as $\omega = \gamma(\alpha \xb) + \tau \fb(\pi_0(\theta))$. Note that the additive form of $\omega$ allows us to separate degree $\gamma$ from type $\tau$. We shall show that the classic mixture version of ZI can be seen as involving parametric but non-linear logistic regression of $\Iy$ on $\xb$. \\

We enter one caveat, illustrated by another single parameter ZI model with somewhat pathological behaviour that is only apparent with close inspection: 
$\omega = \gamma \log(\pi_0)$.  When $\gamma \approx 1$, and except for small $\pi_0$, plots of this function show much in common with the mixture model. 
It too  exhibits the additive property; and it is uniformly inflationary.  But this model implies that 
$\pit_0 = (\gamma -1)\log(\pi_0) - \log(1-\pi_0)$. But when $\pi_0 \to 0$ we find: for $\gamma <1$ that  $\pit_0 \to 0$ - it is not strictly additive; 
and for $\gamma > 1$ we find $\pit_0 \to 1$; that is, $\pit_0(\gamma, \pi_0)$ is not monotone. 
Modelling care may be needed with choosing functions $f(\pi_0)$ and 
data-defined values of $\tau$; there may be algorithmic issues as well, as discussed later.\\

\subsection{Inference and Modelling}

Typically observations are available as $(y_i, \xb_i); i=1,\ldots n$, which we write as $(\yb, X)$. The parameters $(\theta, \omega)$ themselves become parameterised as $\theta_i = \theta(\xb_i \beta)$ and $\omega( \gamma(\alpha\xb_i)) + \tau \fb( \pi_0(\theta_i))$. Inference focusses on the vector parameters $(\alpha, \beta)$ and (if the type of ZI is not pre-specified) on $ \tau$. We proceed via the likelihood in classical inference, and via the likelihood and priors in Bayesian inference. This involves both computation and critical evaluation, the latter involving both model selection and criticism of the data, ideally in a collaboration between statistician and subject matter specialist. \\
 
Computation itself is no longer the main challenge for the size of data sets we consider; but there is natural interest in efficiency, and we turn to this below. The criticism will come most easily from computation that builds on the familiar tools of the statistical trade, which in regression often include graphics and residuals. As we shall see, this typically involves iterating until convergence between: (i) regression of $y$ on $\xb$, for given $\alpha \mbox{ and } \tau$, yielding estimates of $\beta$ and hence of $\pi_{0i} = \pi_0(\theta_i)$; and (ii)  binary regression of $\Iy$ on both $\xb$ and (given $\beta$) suitable functions of  $\pi_{0i}=\pi_0(\theta_i)$, chosen to embody the modelling of the ZI itself, yielding new estimates of $\alpha$ and, if necessary, $\tau$. In particular plots of (estimates of) pairs 
$(\pit_{0i}, \pi_{0i})$ seem likely to be informative. Subject to the (important) qualification that the variance estimates that emerge from these two separate regressions are conditional, not joint, they provide a conventional framework for criticism. We discuss below the details of these regressions and of theory for the appropriate joint inference. \\
    
Initially we explore the theory for \emph{iid} observations. We subsequently consider the regression case where all parameters can be considered as additive 
functions of covariates $\xb$ via coefficients upon which inference becomes focussed.

\section{Maximum likelihood theory for a simple ZI model}

Our interest here lies in likelihood theory for observations $y$ regarded as independent realisations of $\Yt_i \sim \pit_y(\theta, \gamma)$ where $\log \pit_y(\theta, \gamma) = \omega(\gamma, \pi_0)\Iy + \log(\rho) + \pi_y(\theta)$, where $(\theta, \gamma)$ are scalar parameters. We restrict ourselves throughout to the usual case where the mle for $\pi_y(\theta) $  may be studied via  examination of the stationary points of the log likelihood. We first consider  general $\pi_y(\theta)$; but subsequently we focus the case of $\pi_y(\theta) $ in the exponential family. Subsequently we outline the many possibilities beyond this family.\\

In this section we work with \emph{iid} observations. Our focus will be on the special treatment of instances of $y=0$. Of course, with  \emph{iid} data, all two parameter models are equivalent; we note that the essential parameters here are $(\theta,\pit_0)$ and that, here, $\pit_0$ has an obvious estimator ($n_0/n$). Focus therefore rests on the estimation of $\theta$. But the objective is to lay the groundwork for regression, where the choice of ZI model has implications for $\theta$. We concentrate initially on three ZI models: multiplicative, mixture and hurdle. But our interest is wider, for $\omega = \omega(\gamma, \pi_0(\theta))$ provides much flexibility. We subsequently consider the case of regression where $\theta_i = \theta(\xb_i \beta)$ and $\gamma_i = \gamma(\xb_i \alpha)$. \\
 
\subsection{General Properties} \label{Genprops}

We note first some simple properties of $\pit_y( \theta, \omega)$ and $\pi_y(\theta)$. It is simple to show that $\tilde \mu = E[\Yt]=\rho E[Y] = \rho \mu$. More generally, for any function $h(\cdot)$,
\begin{equation*} \label{Eh}
E[h(\Yt)]=h(0)\pit_0 + \rho \sum_{y \ne 0}h(y) \pi_y = h(0)(\pit_0 - \rho \pi_0) + \rho E[h(Y)] = (1-\rho)h(0)+\rho E[h(Y)].
\end{equation*}
Further it is easy to show that, for any functions $h_1(\cdot), h_2(\cdot)
$\begin{equation*}\label{Covh}
Cov[h_1(\Yt)(h_2(\Yt))]= \rho Cov[h_1(Y)(h_2(Y))] +\rho(1-\rho)(E[h_1(Y)]-h_1(0))(E[h_2(Y)]-h_2(0))
\end{equation*}
Hence $Var[\Yt] = \rho Var[Y]+\rho(1-\rho) \mu^2 $.\\

We note also the useful identities below:
\begin{align*}
\ddthfn{ \log(\rho)}  &=  -\ddth \log(1+\kappa \pi_0) 
=-\rho   \left( \kappa \ddthfn{\pi_0}  +\pi_0\ddthfn{(1+\kappa)} \right)\nonumber \\
&=(\rho-1) \ddth \log(\pi_0) -\pit_0 \ddthfn{\omega}; \mbox{ and}\\
\ddomfn{\log(\rho)} &= -\pit_0
\end{align*}
Further we write $\ddthfn{\omega} = \ddpifn{\omega}\ddthfn{\pi_0}=\left(\pi_0 \ddpifn{\omega} \right)\ddth \log(\pi_0)$,  which we write more simply as $u \ddth \log(\pi_0)$.  We also write $v= \ddgmfn{\omega}$ and note that $\ddgmfn{} \log(\rho) = -v \pit_0$. \\

The important term $u = \pi_0 \ddpifn{\omega}$ characterises, for mle, the crucial aspect of the relationship between $\pit_0$ and $\pi_0$, itself the essence of ZI modelling. One useful re-expression involves writing  $g = \logit(\pi_0)$ and $\tilde g = \logit(\pit_0)$; whence $\omega = \tilde g - g$. Then we have an alternative formulation for $u$:
\begin{equation} \label{g_form}
u = \pi_0 \ddpifn{\omega} =  \pi_0 \frac{\partial  \omega }{\partial g}\ddpifn{g}
= \pi_0 \left(\frac{\partial  \tilde g }{\partial g} -1 \right)\frac 1 {\pi_0(1-\pi_0)}
= \left(\frac{\partial  \tilde g }{\partial g} -1  \right)\frac 1 {1-\pi_0}
\end{equation}
The term $\frac{\partial  \tilde g }{\partial g}$ is the slope of the curve in Figure 1B. \\

For the multiplicative model $\omega = \gamma$; thus $u = 0$ and $v = 1$. For the mixture model $\omega = \log(\pi_0 + e^{-\gamma}) -\log(\pi_0 )$; thus $u = - \frac{e^{-\gamma}}{\pi_0+e^{-\gamma} }$, and $v = e^{-\gamma}u$. Recalling that  $\pit_0 = e^{\omega} \rho \pi_0 = \rho(\pi_0+e^{-\gamma})$ for this ZI model, we can re-express this as $u = -\frac{1-\rho}{\pit_0} = -\frac{\kappa}{1+\kappa}$. One implication is that here $\ddthfn{}\log(\rho) = 0$. This, of course, follows more directly from $\rho = q$, independently of $\pi_0$,  which is the distinguishing characteristic of the mixture model. In this it contrasts with the distinguishing characteristic of the multiplicative model, which is that $\omega$ is independent of $\pi_0$. For the hurdle, $\omega =\gamma -\logit(\pi_0)$; thus $u=- \frac 1 {1- \pi_0}$ and $v=1$. For the additive $\omega = \gamma - \log(\pi_0)$; thus $u = -1, v = 1$.

\subsection{Score functions}

We recall that $\log(\pit_y(\theta, \omega)) = \Iy \omega(\gamma, \pi_0) + \log(\rho) + \log(\pi_y( \theta))$. It is useful to denote the score functions \emph{wrt} $\theta$, for $\pit_y$ and $\pi_y$, as $\Stthfn{ y} = \ddth \log(\pit_y(\theta, \omega))$ and $\Sthfn{ y} = \ddth \log(\pi_y( \theta))$. For $\gamma$, we write  $\Stgmfn {y} = \ddgmfn{}\log(\pit_y(\theta, \gamma))$.\\

Then the score functions are:
\begin{align}
\Stthfn{y} &= \ddthfn{}\left(\omega \Iy + \log(\rho) + \log(\pi_y)\right) \nonumber\\
&=(\Iy-\pit_0)u \Sthfn{ 0}+(\rho-1)\Sthfn{ 0} + \Sthfn{ y}, \label{gen_th_mle_eq}\\
\Stgmfn{y} &= (\Iy - \pit_0)v \label{gen_gm_mle_eq}
\end{align}

For the ZI models above, $\Stthfn{y}$ simplifies. For the mixture model, with $u = -\frac{\kappa}{1+\kappa}$, we have 
$$\Stthfn{y}  =(1+\kappa)^{-\Iy}\Sthfn{ y}$$

The implication of this simplification is that $\Sthfn{ y}$ carries a weight of  $(1+\kappa)^{-1} = e^{-\omega}$ when $y=0$. When $\kappa>0$ this weight is $P(Y=0 |\Yt=0)$, in the conventional notation of this ZI model, and one important component of Lambert's EM algorithm. But note that the multiplier has value 
$(1+\kappa)^{-1}$ for \emph{all} $\kappa$, and  can always can be interpreted as the expected number of instances of $Y=0$ associated with each $\Yt=0$, which exceeds one for $\omega < 0$, that is, for under-inflation. Recall that under-inflation in this model must be constrained to $\pit_0 >0$, and is thus subject to  subject to  $\gamma < -\log(\pi_0)$.\\

For the hurdle, recalling that $\rho = \frac{1-\pit_0}{1-\pi_0}$, we find that 
\begin{align*}
\Stthfn{y} &= - \frac{1-\pit_0}{1-\pi_0}\Sthfn{ 0}  +(\rho-1) \Sthfn{ 0}+ \Sthfn{ 0} = 0 \mbox{, for } y =0 \mbox{ and}\\
&= \frac{\pi_0}{1-\pi_0}\Sthfn{ 0} +  \Sthfn{y} \mbox{, for } y > 0
\end{align*}
The implication of $\Stthfn{0} =0$ is that here instances of $\Yt=0$ do not contribute to inference on $\theta$, as is standard in the hurdle. It is easy to establish that $\Stthfn{y} = \ddth \log (\pi^+_y(\theta))$, where $\pi^+_y(\theta)= \frac{\pi_y}{1-\pi_0}$ is the truncated version of $\pi_y(\theta)$.\\

For the multiplicative, with $u= 0$, we have $\Stthfn{y} = (\rho-1)\Sthfn{ 0} + \Sthfn{ y}$. We will see below that this  is particularly simple for $\pi_y$ in the exponential family. 
\\

Compactly we may write, with $m_Y=\Sthfn{ 0}$,
\begin{equation} \label{gen_score}
\tilde S(y)=\left[\begin{array}{c}\Stth(y)  \\ \Stgmfn{y}\end{array}\right]
=G\left[\begin{array}{c}\Sthfn{y}  \\ \Iy -\pit_0 \end{array}\right]
+\left[\begin{array}{c}(\rho-1)m_Y  \\ 0 \end{array}\right]
\end{equation}
where $G = \left(\begin{array}{cc}1 & u\Sthfn{0} \\0 & v\end{array}\right)$. We shall see that, when $\pi_y(\theta)$ is in the exponential family,  $m_Y = E[Y]$.\\

\subsection{Information Matrices}

Fisher's expected information, $\FIt = Var\left[\begin{array}{c}\Stth(\Yt)  \\ \Stgmfn{\Yt}\end{array}\right] $ follows directly from \eqref{gen_score} via the consideration of $Var\left[\begin{array}{c}\Sthfn{\Yt}  \\ \IYt -\pit_0 \end{array}\right]$. From Section \ref{Genprops} we have
$$Var\left[ \Stthfn{\Yt}\right]=\rho Var\left[ \Sthfn{Y}\right] +\rho(1-\rho)m_Y^2.$$
Clearly $Cov\left[ \Sthfn{Y}, \IY \right] = m_Y$ and $Var\left[ \IYt\right]=\pit_0(1-\pit_0)$. Then $\FIt = G H G^T$, where 
$$H = \left(\begin{array}{cc}\rho Var\left[ \Sthfn{Y}\right] +\rho(1-\rho)m_Y^2 & m_Y \\
m_Y & \pit_0(1-\pit_0)\end{array}\right)$$.\\

The observed information matrix is available, via additional identities for $\ddthfn{\rho} $ and $\ddthfn{\pit_0}$ which follow from $\ddthfn{\rho} =\rho \ddthfn{}\log(\rho) $ and $\ddthfn{\pit_0} = \pit_0 \ddthfn{}\log(\pit_0)$ and from the second derivatives of $\omega(\gamma, \pi_0)$. For the mixture model these second derivatives are not insightful. But for simpler cases where $\omega = \gamma + \tau\fb(\pi_0)$ only the second derivative \emph{wrt} $\theta$ are non-zero, leading to simplifications.\\

The information matrix provides access to an approximation to the \emph{joint} variance of the estimators $(\hat \theta , \hat \gamma )$.

\subsection {The case of \emph{iid} data for general $\pi_y(\theta)$} 

Our focus on the \emph{iid} case comes via the inferential role of the instances of $y_i=0$.  Of course here all two parameter models are equivalent, for all may be characterised by $(\theta, \pit_0)$.  However, the expected information is not the same for all ZI models, reflecting the fact that the parameter $\gamma$ has different interpretations in different models.\\

Given  $n$ \emph{iid} observations $\yb$, of which $n_0$ have value zero the mle's $(\hat \theta, \hat \gamma)$ satisfy:
$\Stthfn{\yb}  = \sum \Stthfn{y_i}=0$ and $\Stgmfn {\yb}  = \sum  \Stgmfn {y_i} = 0$ where
\begin{align*}
\Stthfn{\yb}  &=\left( \sum(\Iyi - \pit_0)\right)u\Sthfn{0} 
+ n(\rho-1)\Sthfn{0}+\sum_i \Sthfn{ y_i} \\
\Stgmfn {\yb}  &= \left(\sum_i \Iyi - n \pit_0\right)v = (n_0 -n \pit_0)v 
\end{align*}
 
From the latter, $\hat {\pit}_0 = \pit_0(\hat \theta, \hat \gamma) =  \frac{n_0} n$, a trivial result in the \emph{iid} case. The former then simplifies to estimators, including $\hat \theta, \hat\pi_0=\pi_0(\hat \theta),  \hat \omega= \omega(\hat \gamma, \hat\pi_0) = \log(1+\hat \kappa)$ and similarly $ \hat \rho = \frac{1-\hat {\pit}_0}{1-\hat \pi_0}$, being such that they satisfy
$$ \Stthfn{\yb}  =\left(n(\hat \rho-1)+n_0 \right) \Sthfn{0} + \sum_{y_i>0} \Sthfn{\theta; y_i} $$
The terms $(u,v)$ have cancelled in this \emph{iid} case. Thus confirms, of course, that here all two-parameter ZI models have the same mle's $(\hat \theta, \hat{\pit}_0)$. \\
 
\subsection{$\pi_y$ in the exponential family}

Alternative forms of the score equations arise for $\pi_y(\theta)$ from within the exponential family; that is $\log(\pi_y(\theta))= \theta y - A(\theta)$ to within an additive function of $y$. For then we have: 
\begin{equation*} \label{ExpFam}
\log(\pit_y( \theta, \omega))=\Iy\omega + \theta y + \log(\rho) -A(\theta)  = \Iy\omega + \theta y  -\tilde A(\omega,\theta) 
\end{equation*}
with $\tilde A(\omega,\theta) = A(\theta) - \log(\rho)$.  \\
 
Now, when $\omega(\gamma, \pi_0)=\gamma$ is independent of $\pi_0$, then in this context $\pit(\theta, \gamma)$ is a member of the  two parameter exponential family with natural parameters $( \theta,  \gamma)$. This confirms, in a precise sense, the earlier statement that multiplicative ZI is the simplest case. Now $\Stthfn{ y} = y-\rho \mu$ (with $\mu = A^{\prime}(\theta)$) may be interpreted as $\Stthfn{ y} = y-E[\Yt]$; similarly $\Stgmfn{ y} = \Iy-E[\IYt]$.  In the \emph{iid} case the sufficient statistics are $n_0 = \sum \Iyi$ and $\sum y_i$. Note that the sample mean $\bar y$ is the mle for $E[\Yt] = \rho \mu(\theta)$. Given $\rho$, the rescaled sample mean is unbiased for $\mu$. But $\rho$ is not known.\\

For a general ZI model Equations \eqref{gen_th_mle_eq} and \eqref{gen_gm_mle_eq} are now written more simply as
\begin{align}
\Stthfn{y} &= - (\Iy - \pit_0)u\mu + (y-\rho \mu) \label{EF_th_mle_eq}\\
\Stgmfn{y} &= (\Iy - \pit_0)v \label{EF_gm_mle_eq}
\end{align}

The expected information is 
\begin{align} \label{FishInf}
\FIt &= \left(\begin{array}{cc}1 & -u \\0 & v\end{array}\right) 
Var\left[\begin{array}{c} \Yt \\ \IYt\end{array}\right]
\left(\begin{array}{cc}1 & 0 \\-u & v\end{array}\right) \nonumber\\
&= \left(\begin{array}{cc}1 & -u \\0 & v\end{array}\right) 
\left(\begin{array}{cc}\rho Var[Y] +\rho(1-\rho)\mu^2 & -\rho\mu\pit_0 \\-\rho\mu\pit_0 & \pit_0(1-\pit_0)\end{array}\right)
\left(\begin{array}{cc}1 & 0 \\-u & v\end{array}\right) 
\end{align}
since $Cov[\Yt, \IYt] = E[\Yt \; \IYt]-E[\Yt]E[\IYt] = -\rho\mu\pit_0$ as $\Yt\; \IYt=0$. This is particularly simple when $u=0$ for the multiplicative ZI.

\subsection{The case of \emph{iid} data for $\pi_y(\theta)$ in the exponential family}

The score equations are $\sum_i \Stthfn{y_i} = 0$ and $\sum_i \Stgmfn{y_i} = 0$. As previously the terms $(u,v)$  cancel, and $ \hat {\pit}_0 = \frac{n_0} n$. But now for $\pi_y$ in the exponential family,  $\hat \theta$ is the solution to $\sum(y_i-\rho \mu) = 0$. \\

For multiplicative ZI, $u=0$, whence $\phi = \rho, \psi =1$: thus  $\Stthfn{y} = y -\rho \mu$. Thus for multiplicative ZI, $(\hat \theta, \hat \gamma)$, or equivalently for $(\hat \theta, \hat \rho)$, are the solution to the equations $\sum_i (y_i-\rho \mu(\theta)) = 0$ and $\sum_i (\Iyi- \pit_0) = 0$. A natural iterative algorithm  begins  with $ \hat \theta^{(1)} =  \bar y$ from an initial  $ \hat \rho^{(0)} = 1$; then with $ \hat {\pit}_0 = \frac{n_0}{n}$ and $ \hat \rho^{(1)} = \frac{1-  \hat {\pit}_0}{1-\hat \pi_0}$, we have $\hat \theta^{(2)} = \left(\hat \rho^{(1)}\right)^{-1}\bar y$; repeating until convergence. We refer to this as an iterative re-scaling algorithm. \\
 
Observe that there is no special treatment of zeroes, as arose with the mixture and hurdle formulations of ZI; yet 
for the \emph{iid} case these solutions must lead to the same estimators.  It is insightful, both here and  for the later purpose of studying regression, to re-express \eqref{EF_th_mle_eq} as 
\begin{equation} \label{wted_mle}
\Stthfn{y} = -\Iy u \mu  +  y - (\rho - u\pit_0)\mu = \psi^{\Iy}(y-\phi \mu)
\end{equation}
where $\phi = \rho - u\pit_0$ and $\psi = \frac{u+\phi}{\phi} =1 + \frac{u}{\phi}$. The term $\psi$ is a special weight on instances of $y_i=0$. 
\\

Thus we can write the score equation for $\theta$ in a general form, which will be important in regression.
\begin{equation} \label{EstEq}
\Stthfn{\yb}  = \sum_i \psi^{\Iyi}(y_i-\phi \mu) = 0
\end{equation}
Then the solution must satisfy the weighted average $\phi \mu = \frac{\sum \psi^{\Iyi}y_i}{\sum \psi^{\Iyi}}= \frac{\sum {y_i} } {\sum \psi^{\Iyi}}$. Here the term $n^* = \sum \psi^{\Iyi}$ may be thought of as an effective sample size (ESS); this formulation thus generalises \citet{cohen1960}. All ZI models, and thus all $(\phi, \psi)$ terms,  lead here to the same ESS as we outline. \\

It is useful to consider the mixture version of Equation \eqref{EstEq}. This solves:
\begin{equation} \label{wted_mle}
\Stthfn{y} = \sum (1+\kappa)^{-\Iyi}(y_i - \mu) = 0
\end{equation}
The natural algorithm model now (in the case of $\pi_y(\theta)$ in the exponential family) leads to $\hat \theta$  being the solution obtained by iterative re-weighting of instances of $y=0$. That is, from an initial $\hat \kappa^{(0)} = 0$ we obtain $\hat \theta^{(1)}$ satisfying $\mu(\theta) = \bar y$; the equation $\pit_0 = \frac{n_0} n$ leads to $\hat \kappa^{(1)}$; the iteration proceeds to convergence. As we shall see, in the wider context of regression where the ZI models lead to different solutions, the natural algorithm solves equations based on \eqref{wted_mle} by iteratively re-scaling and re-weighting. In the \emph{iid} case, the exponential family formulation leads to all versions having the same ESS.\\

However, there is a subtle theoretical point for $n_0$ that can be seen as a realisation of $N_0 = \sum \Iyi$. Then the expected value of the (random) effective sample size $N^*$ is $E[N^*] = E\left[ \frac{N_0}{1+\kappa} + (n- N_0)\right] = n\rho$. But $E\left[ \hat \theta \right] = E\left[\frac{\sum_i \Yt_i}{N^*} \right] \ne \frac{\sum_i E[\Yt_i]}{E[N^*]} = \frac {n\rho\mu}{n\rho} = \mu$. Thus, unlike the re-scaling estimate with known $\rho$, the weighted estimate with known $\kappa$ is not unbiased for $\mu$. It is of course asymptotically unbiased. In the special case of \emph{iid} data and unknown parameters the estimators are identical, but both are biased. \\

\subsubsection{$\phi$ and $\psi$ are non-negative}
It is useful to know, for the purposes of algorithms,  that for all $\pi_y(\theta)$ in the exponential family the both of the weights $\phi$ and $\psi$ are non-negative, the latter subject to the important constraint that $\pit_0$ is a non-decreasing function of $\pi_0$. We demonstrate below by exploring the implications of 
$\phi=0$ and $\psi=0$. \\

Starting from \eqref{g_form}:
$$\phi = \rho -\pit_0 u = \frac{1-\pit_0}{1-\pi_0} +\frac{\pit_0}{1-\pi_0}  \left(1 - \frac{\partial  \tilde g }{\partial g} \right)
=  \frac{1}{1-\pi_0}\left( 1- \pit_0 \frac{\partial  \tilde g }{\partial g} \right)$$
Thus $\phi \ge 0$ when $\frac{\partial  \tilde g }{\partial g} \le \frac 1 {\pit_0}$ or, equivalently, when $\frac{\partial   g }{\partial \tilde g} \ge \pit_0 = \frac{e^{\tilde g}}{1+e^{\tilde g}}$. But $\phi=0$ leads to $\frac{\partial   g }{\partial \tilde g}=\frac{e^{\tilde g}}{1+e^{\tilde g}}$, which requires that $g = \gamma + \log(1+e^{\tilde g}) = \gamma -\log(1-\pit_0)$ for some constant $\gamma$; or, equivalently, when $\pit_0 = 1- e^{\gamma}\frac{1-\pi_0}{\pi_0} = (1+e^{\gamma}) -\frac{e^{\gamma}}{\pi_0}$, providing, of course, that $0 \le \pit_0 \le 1$. But this in turn requires that $\frac{e^{\gamma}}{1+e^{\gamma}} \le \pi_0 \le 1$; that is, $\pi_0$ is bounded below. However, this is contradicted by $\pi_y$ being in the exponential family unless $e^{\gamma} = 0$, and this in turn requires that the degenerate case of $\pit_0 = 1$ is the only model for which $\phi = 0$. We thus assert that, apart from this degenerate case, $\phi > 0$ for  $\pi_y$ in the exponential family.\\

As regards $\psi$, we  note that $\psi =  \frac{u+\phi}{\phi}$ and focus is on the sign of 
$$u+\phi = \rho+(1-\pit_0)u = \rho + \frac{1-\pit_0}{1-\pi_0}\left( \frac{\partial  \tilde g }{\partial g} -1 \right) =\rho \frac{\partial  \tilde g }{\partial g}.$$
Thus if $\frac{\partial  \tilde g }{\partial g} \ge 0$, that is if $\pit_0$ is an increasing function of $\pi_0$, then $\psi \ge 0$, since $\phi \ge 0$. Note that $\frac{\partial  \tilde g }{\partial g} = 0$ characterises the hurdle model.\\

We note that the attractive but possibly pathological case of $\omega = \gamma \log(\pi_0)$ has $u = \gamma$ and $v = \log(\pi_0)$. But recall that 
when $\gamma > 1$, $\pit_0$ is a \emph{decreasing} function of $\pi_0$, for small $\pi_0$. It follows that the special weight $\psi$, on instances of $y_i=0$
 can be negative. 
\\

We now turn to  the wider context of ZI regression. Here  ZI is modelled by arbitrary functions $\omega(\gamma_i,\pi_{0i})$, where $\gamma_i = \gamma(\xb_i \alpha)$ and $\pi_{0}(\theta(\xb_i \alpha))$, the key new issue is that we no longer have the simple $\hat {\pit}_0 = \frac {n_0}{n}$. The estimation of $\alpha$ is the essential new issue; this is ZI modelling. The estimation of $\beta$ is that of regression in the presence of $ZI$.\\

\section{Regression models for $\theta$ and $\gamma$ in the exponential family}

We now consider maximum likelihood for the case where the elements $y_i$ of $\yb$ are taken as independent realisations of $\Yt_i \sim \pit_y(\theta_i,\gamma_i)$ where $\theta_i=\xb_i\beta$ and $\gamma_i=\xb_i\alpha$ in the context of  $\omega(\gamma, \pi_0)= \omega( \gamma(\alpha\xb)) + \tau \fb( \pi_0(\theta))$. We initially consider the ZI type, and thus assume that the coefficients $\tau$ and the functions $\fb(\cdot)$, to be known. We focus exclusively on the case of the exponential family. We consider separately the estimation of $\beta$ for given $\alpha$ and of $\alpha$ for given $\beta$, envisaging an algorithm that iterates between these from an initial $\alpha$ corresponding to the absence of ZI. The former builds on the \emph{iid} estimating equation at \eqref{wted_mle}; the latter generalises the trivial result for the \emph{iid} case, $\hat {\pit}_0 = \frac{n_0} n$, and is thus the main focus of this section. Initially we focus on the case where the ZI model is itself known; that is, the coefficients $\tau$ and the functions $\fb(\pi_0)$ are known. Subsequently we consider the modelling of ZI itself.

\subsection{$\beta$ parameters, for given $\alpha$} 

The score equations for each $\beta_j$ follow directly from \eqref{EF_th_mle_eq}. It follows that, for case $(y_i, \xb_i)$,
\begin{equation}
\Stbetafn{y_i} = \grad_{\beta} \log(\pit_{y_i}(\theta_i, \omega_i)) = \Stthfn{y_i} \grad_{\beta}(\theta_i) = \psi_i^{\Iyi}(y_i-\phi_i \mu(\theta_i))\xb_i
\end{equation}
where $\omega_i = \omega(\gamma_i, \pi_0(\theta_i))$ and $\gamma_i = \alpha \xb_i$, and $(\psi_i, \phi_i)$ are defined by these. Thus
\begin{equation} \label{betafit}
 \Stbetafn{\yb} =\sum_i  \psi_i^{\Iyi}(y_i-\phi_i \mu(\theta_i))\xb_i = \sum_i  \left( \psi_i^{\Iyi} \phi_i \right)(\dot{y}_i-\mu(\theta_i))\xb_i
\end{equation}
where $\dot{y}_i = \phi_i^{-1}y_i$. In the second form we recognise, for given $(\psi_i, \phi_i)$, the estimating equations for  a weighted quasi-(log)likelihood, the weights being $\psi_i^{\Iyi} \phi_i$, the likelihood being based on that of $\pi_y(\theta)$;  see for example \citet{pawitan2001} Ch 14. Note that $\dot{Y}=  \phi_i^{-1}Y$ is no longer integer; $\pi_y(\theta)$ is not its pmf. One solution is thus via iterative, weighted, quasi-loglikelihood, the weights depending on $\gamma_i$ and thus on $\alpha$. Ultimately, such an algorithm is iteratively re-weighted least squares. But typically the coefficients $\alpha$ are themselves not known. \\

\subsection{$\alpha$ parameters, for given $\theta_i$ } 

The estimation of the $\alpha$ parameters follow from score equations built on \eqref{EF_gm_mle_eq}; but here the terms
$v_i =  \ddgmfn{\omega_i}$ no longer cancel. Formally the mle for $\alpha$ satisfies
\begin{equation} \label{alphafit}
\Stalphafn{\yb} =\sum_i  ({\Iyi} - \pit_0(\theta_i))\xb_i
\end{equation}
It is useful to interpret this in the light of Figure 1. We consider first the simplest case, that of Multiplicative ZI with $\omega = \gamma$ being a scalar independent of $\xb_i$ and thus common to all cases. Then an estimate of $\gamma$ is returned, as the intercept, by regular linear logistic regression of $\Iyi$ on (known) 
$\logit (\pi_0(\theta_i))$, in which the coefficient of $\logit (\pi_0(\theta_i))$ is forced to unit value. The fitted values are estimates of $\pit_0(\gamma, \theta_i)$ with here the common $\gamma$ being estimated by the intercept. This is most simply achieved by using $\logit (\pi_0(\theta_i))$ as an offset term in a linear logistic regression of $\Iyi$ on the (common) unit vector.  For this ZI model there is a linear relationship, with unit slope,  between fitted $\logit (\pit_0(\theta_i))$ and $\logit (\pi_0(\theta_i))$ as in Figure 1. The algorithmic solution of \eqref{alphafit} is thus straightforward, for given fitted $\beta$. The joint estimation of $(\alpha, \beta)$ thus involves iterating between \eqref{betafit} and \eqref{alphafit} until convergence.\\

The extension to $\omega_i = \alpha \xb_i$ is similarly achieved by logistic regression of the $\Iyi$ on covariates $\xb_i$, again with an offset. Of course, if there is significant variation on $\gamma_i$ the plot of $\logit(\pit_{0i})$ against $\logit(\pi_{0i})$ will not be as simple as in Figure 1. Nevertheless, the routine diagnostics of linear logistic regression will be of some assistance in criticising this choice of ZI model, albeit that without modification, these are strictly interpretable on in the sense that it is conditional on  $\theta_i$ being known. We outline below the route to the joint inference, and thus to the modification.\\

The case of the hurdle model of ZI corresponds, in the simplest case, to the logistic regression of $\Iyi$ on the unit value. The fitted values of $\logit (\pit_0(\theta_i))$ will thus be independent of $\logit (\pi_0(\theta_i))$. The extension to $\omega_i = \alpha \xb_i$ is as in the simplest case. The  additive model of ZI involves a regression on $\xb_i$ and $\log(1-\pi_0(\theta_i))$, with the coefficient of $\log(1-\pi_0(\theta_i))$ forced to value -1, again achieved by offset. More generally, when the ZI model is specified by known values $\tau$ on known functions $f(\pi_0(\theta_i))$, then these form the offset in a regression of $\Iyi$ on the covariates, subject to the caveat illustrated above by the case of $\omega = \gamma \log(\pi_0)$. \\

The mixture model is more challenging algorithmically. But it too may be seen as logistic regression of $\Iyi$ on the non-linear function 
$\logit ((1-q)+q\pi_0(\theta_i))$ of $\pi_0(\theta_i)$, which is most naturally parameterised as $\logit \left(\frac{1+e^{\gamma_i}\pi_0(\theta_i)}{1+e^{\gamma_i}}\right)$ where $\gamma_i = \alpha \xb_i$. The real imagination in Lambert's EM is the realisation that this can be achieved by conventional, albeit weighted, linear logistic regression. For, given an estimate of $\beta$ and thus values of $\pi_0(\theta_i)$, it is possible to compute the equivalent of $\frac 1 {1+\kappa_i} = e^{-\omega_i}$ for each case. When $\kappa_i >0$, this can be interpreted as the case specific value of $Pr(J_i=1 | \Yt_i=0)$ for each of the $n_0$ instances of $y_i=0$ in the data set. Of course, $Pr(J_i=1 | \Yt_i=y_i) = 1$ for each of the $(n-n_0)$ instances of $y_i >0$.\\

Lambert's  EM algorithm logistically regresses (latent) $J$ against covariates $\xb$ by the device of logistically regressing, with weights, $n_0+n$ values of $J$ on corresponding $\xb$ as follows: for each of the $n-n_0$ cases with $y_i> 0$, create a case with $J_i =1$, associating it with its covariate vector $\xb_i$ and unit weight; for each of the $n_0$ cases with $y_i=0$ create two sets of cases: $n_0$ cases with $J_i=1$,  corresponding $\xb_i$  and weight $\frac{1}{1+\kappa_i}= e^{-\omega_i}$; and a further set of $n_0$ cases with $J_i =0$, the same corresponding $\xb_i$  but now weight $1- e^{-\omega_i}$; we refer to this 
artificial data set as $(J^+, X^+)$. The weighted logistic regression of these $n+n_0$ values of $J^+$ against their corresponding $\xb^+$ yields the mle for coefficients $\alpha$. Although this is efficient, it is somewhat opaque to critical evaluation of the mixture model for ZI.  Irrespective of  which algorithm is used to fit the unknown $\alpha$, plots of $n$ pairs  
$\left( \hat {\pit}_{0i}, \hat {\pi}_{0i}\right)$ in either the linear or logistic metrics (as in Figure 1) may be helpful, here as with all ZI models.\\

\subsection{Expected information}
The Expected information provides access to joint inference. In the case of the multiplicative model it is exact, for here $\pit_y$ is in the two-parameter exponential family. It comes directly from \eqref{FishInf}. Denoting as row-vectors $\xb_i^{\beta}$ and $\xb_i^{\alpha}$ the subsets  of covariates involved modelling $\theta_i$ and $\gamma_i$ we find
\begin{equation}
\FItfn {\beta, \alpha} = \sum_i \left(\begin{array}{c}\xb_i^{\beta} \\ \xb_i^{\alpha}	\end{array}\right)^T
							\FItfn{\theta_i,\gamma_i}
							\left(\begin{array}{c}\xb_i^{\beta} \\ \xb_i^{\alpha}	
							\end{array}\right)\\
\end{equation}
Note that the subsets $\xb_i^{\beta}$ and $\xb_i^{\alpha}$ may in principle overlap, contrary to a restriction imposed by the EM algorithm when used to fit the Mixture model. It is to be anticipated that, as always but especially here, multi-collinearity will be encountered if the model is over-parameterised. But the options for avoiding this are increased by the availability of multiple different models for the ZI aspect of the model.\\

\subsection{Choice of ZI model}
If the choice of ZI model can be restricted to that which can be written as $\omega =\gamma + \sum_k \tau_k f_k(\pi_0)$, with $\gamma =  \alpha \xb$, then this is a simple extension of the above linear logistic regression. For now we may regress $\Iyi$ against both $\xb_i$ and the $f_k(\pi_{0i})$, estimating the coefficients $\tau_k$, and thus the ZI model. \\

Arguably, this may be the simplest way to choose the ZI model, especially if flexibility in the $\tau$ coefficients permits a parsimonious model for $\gamma$. Conventional diagnostics may provide access to some criticism of the chosen model; for then the resultant plot of (the fitted values of) $\pit_{0i}$ vs $\pi_{0i}$ may provide the simplest way to present the ZI model, which is no longer required to take one of a pre-chosen suite of models. Recall that current practice at best offers a comparison of the two extant ZI models (mixture and hurdle), typically in terms of a portmanteau measure such as BIC.\\

Indeed this may be enriched further to cater for the possibility that the ZI model is itself a function of the covariates $\xb$; for this corresponds to allowing for interaction between the covariates and the functions $f_k(\pi_0)$. It is to be anticipated however that only data sets with a very strong signal to noise ratio will have the power to allow discrimination between alternative ZI models; and even fewer will allow meaningful identification of interactions. The caveat on non-decreasing ZI models may also constrain the search. \\

In closing we draw attention to other possibilities. It is not strictly necessary that the binary regression, with offset of $\logit(\pi_0)$, be \emph{logistic} regression; for example $\mbox{probit}(\cdot)$ or other functions may be used to map from $\Re$ to the interval (0,1). For example, the model used by \citet{ghosh2006bayesian}, in addressing problems with very large numbers of observed zeroes, may be thought of as using the cdf of the beta-binomial distribution. There is a need howerer for careful consideration of the behaviour as $\pi_0 \to 0$ and $\pi_0 \to 1$.

\section{Examples}

We illustrate the power of this new approach with a simple example illustrated through Bayesian Inference via the Hamiltonian Monte Carlo package \texttt{rstan} \citep{rstan2018}. We use a simple zero-inflated fishing data set to illustrate the use of model comparison techniques to determine the type of zero inflation amongst distinct sub-groups according to a covariate. The example is further illuminated by the plot of $(\pi_0, \pit_0)$ which displays the type of ZI behaviour. Code to run the examples is available at www.github.com/andrewcparnell/ZIpaper.\\

\begin{figure}[!h]
\includegraphics[width = \textwidth]{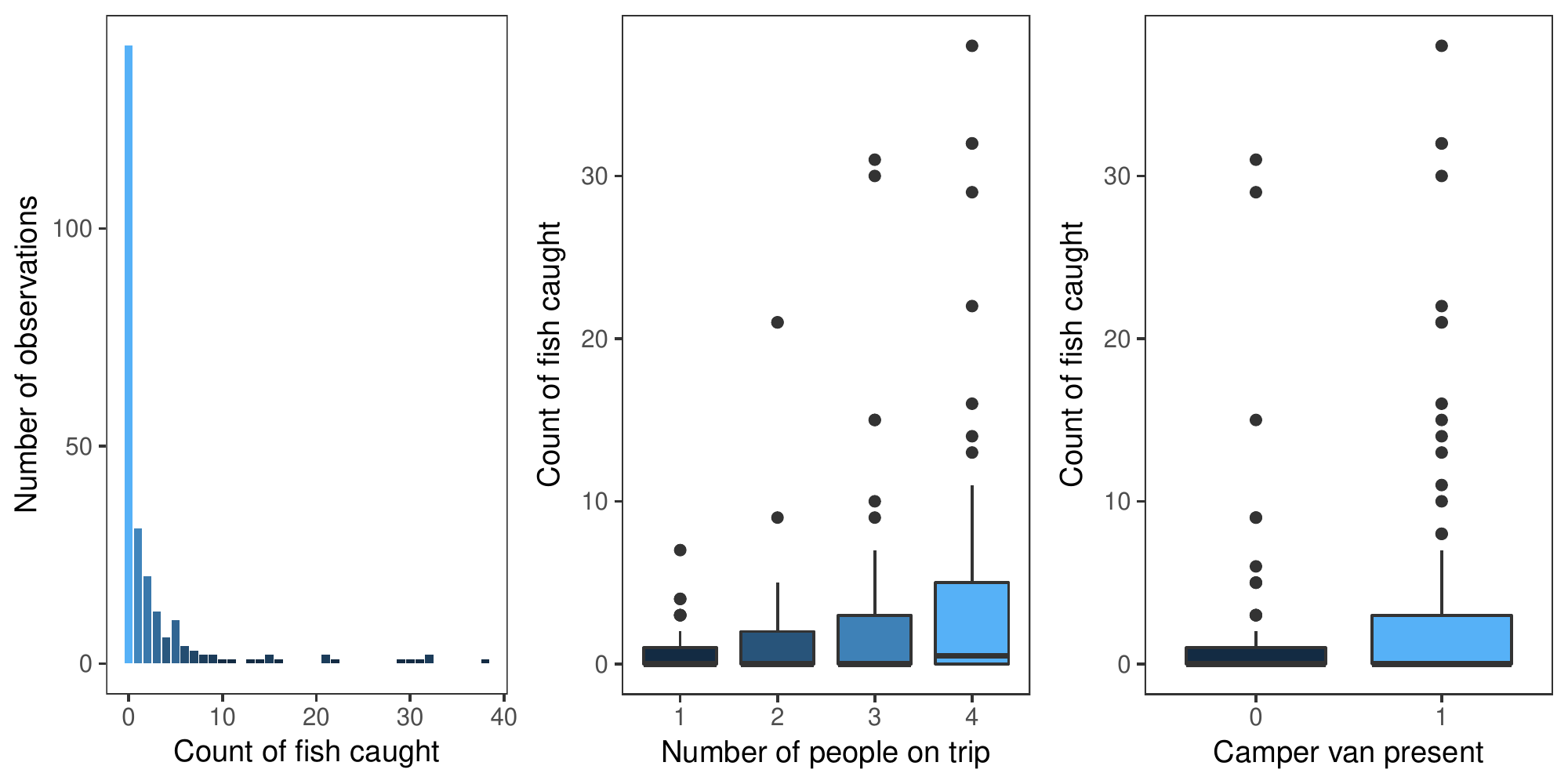}
\caption{A plot of the fish data set. The left panel shows a histogram of the response, the middle panel a boxplot of the response against the covariate \texttt{persons}, and the right panel a boxplot of the response against the covariate \texttt{camper}.}
\label{fig:fish1}
\end{figure}

We use a data set from the UCLA Academic Technology Services website (http://
www.ats.ucla.edu) and previously analysed in \citet{saffari2013investigating}. The data concern an environmental survey by a state wildlife biologist of groups that went to a park and attempted to catch fish. The number of fish they caught was recorded (variable \texttt{count}), as was further data on each group. We use the additional variables \texttt{persons} (number of people attending) and \texttt{camper} (whether or not they brought a camper van). Two counts had excessively high values of 65 and 149 and are removed from this exploratory analysis. The count response we write as $y_i, i = 1, \ldots, n$ and covariates $\xb_1, \xb_2, \ldots, \xb_n$ where $n=248$. A histogram of the response and boxplots of the main covariates are shown in Figure \ref{fig:fish1}.\\

We define the new distribution as $ZIPo$ with pmf:
\begin{equation*}\label{basic}
\pit_y(\kappa, \theta) = \left\{ \begin{array}{ll} (1+\kappa)\rho \pi_0(\theta) & \mbox{if } y = 0 \\ \rho \pi_y(\theta) & \mbox{otherwise}. \end{array} \right.
\end{equation*}
with $\omega = \log(1 + \kappa)$ and $\rho = (1 + \kappa \pi_0)^{-1}$.\\

For simplicity, we use the Poisson as the base distribution. The model can be written out hierarchically as: 
\begin{eqnarray*}
y_i &\sim& ZIPo(\omega_i, \theta_i) \\
\omega_i &=& \alpha  - \tau_1 \theta_i + \tau_2 \log(1- e^{-\theta_i})\\
\log(\theta_i) &=& \beta \xb_i\\
\alpha &\sim& N(0, 10^2) \\
\beta_{k} &\sim& N(0, 10^2)
\end{eqnarray*}
where $k$ is the covariate number and vague priors are given to the hyper-parameters. The different ZI models can be found through the values of $\tau$. They are $(\tau_1, \tau_2) = (0,0), (0,-1), (-1,-1)$ for the multiplicative, additive, and hurdle respectively. We identify these by fitting the model once for each type, and subsequently calculating the WAIC values \citep{Watanabe2013, Gelman2013}.\\

\begin{figure}[!h]
\includegraphics[width = \textwidth]{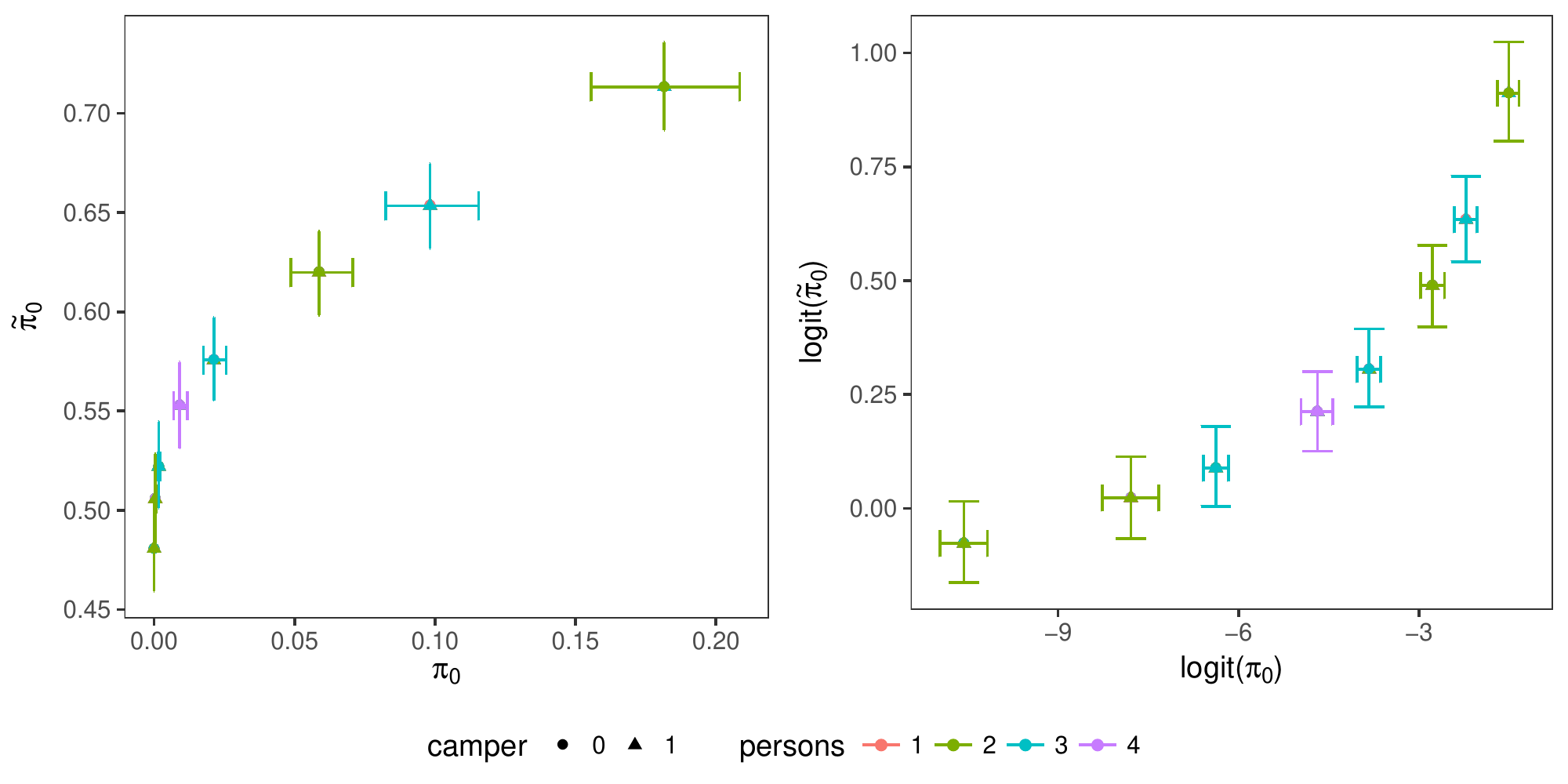}
\caption{A plot of the posterior medians of $\pi_0$ vs $\pit_0$ (left panel) and the same on the logit scale on the right panel. This plot should be contrasted with Figure 1 to identify behaviour. The different covariate values used for the plot are shown in the legend.}
\label{fig:fish2}
\end{figure}

We run the models for the default 2000 iterations across 4 chains with 1000 as burn-in and check convergence using the standard R-hat diagnostic \citep{brooks2011handbook}. We subsequently compute WAIC values using the \texttt{loo} package \citep{vehtari2016}. Table \ref{tab:fish1} shows the results. The hurdle model seems to be strongly selected over the others. However we can extend this model to include situations where different ZI approaches may be preferred by different covariate values. The plot of $(\pi_0$ vs $\pit_0$ is shown in Figure \ref{fig:fish2} which, most interestingly, shows some elements of non-Hurdle like behaviour. 

\begin{table}[!h]
\begin{tabular}{lrr}
\hline
ZI type & WAIC & se(WAIC) \\
\hline 
Hurdle & 1,277 & 143\\
Multiplicative & 1,359 & 159\\
Additive & 1,365 & 159\\
\hline
\end{tabular}
\caption{WAIC values with standard errors for a simple new ZI model applied to the fish data}
\label{tab:fish1}
\end{table}

\section{Extensions}

We see multiple extensions. These include: Bayesian inference with latent processes; the use of other families of univariate count distributions; multiple inflations of count data; extensions to continuous distributions; and multivariate zero-inflations of count data. We outline some options.

\subsection{Bayesian inference with latent processes}

With $(\theta, \gamma) = \left( \theta(\xb \beta),\gamma(\xb \alpha)\right)$ we seek the posterior conditional distribution $[\beta, \alpha | \Ybt = \yb ]$ which for simplicity of notation, we write in this section as $[\beta, \alpha | \Ybt]$. In is natural to approach this by sampling from $[\beta, \alpha|  \Ybt, \IYbt, X] $, Gibbs fashion, by successively sampling from the full conditionals: 
$$[\beta | \Ybt, \IYbt, \alpha,X] = [\beta| \alpha, \Ybt,X]   \hspace{1cm} \mbox{ and  } [\alpha | \Ybt, \IYbt, \pi_0(\theta),X] = [\alpha | \IYbt, \pi_0(\theta),X]. $$
The latter indicates the usual sampling in classical Bayesian inference for logistic (or more general binary) regression of observed instances of $\IYt$ on suitable functions of $\pi_0(\theta)$. We thus presume that $[\alpha | \IYbt, \pi_0(\theta),X]$ is available and focus therefore on the former. \\

We shall presume the existence of an algorithm to sample from $[\beta | \Yb = \yb,X]$, that is, Bayesian inference in the regression of count data, absent ZI,  on covariates $\xb$. Here we outline how to modify $[\beta | \Yb,X]$, in the presence of a known type of  ZI, to build an algorithm to sample from 
$[\beta | \Ybt, \alpha,X]$. We then propose the use of this algorithm to sample from $[\beta, \alpha | \Ybt,X]$ above.  \\
  
We recall that we can associate, with every $\Yt =y$, a latent integer $M$, denoting the implicit number of observations of $Y=y$. Clearly when $y\ne 0$ then $M=1$. But implicit in our model of ZI, is that when $y=0$, $M=1$ is only one possibility. In particular, with over-inflation, $M$ is binary, with 
$P(M=1 | \Yt=0) = E[M |\Yt=0] = \frac 1 {1+\kappa} = e^{-\omega}$; with under-inflation, however, $M$ has a geometric distribution on $\ZZ_0^+$ with $E[M | \Yt=0] =\frac 1 {1+\kappa} = e^{-\omega}$. We note also that 
$$[\beta | \Ybt, \alpha,X] = E_{\Mb | \Ybt, \alpha}\left[  \beta  | \alpha, \Mb, \Ybt,X \right] 
= E_{\Mb | \IYbt, \alpha}\left[  \beta  | \alpha, \Mb, \Ybt, X\right].$$
But the joint knowledge that $\Yt_i = y_i$ and $M_i=m_i$ tells us that we can treat $y_i$, for inferential purposes, as $m_i$ copies of $Y_i=y_i$ and of its covariates $\xb_i$; we write this as $\left(y^M_i,\xb^M_i\right)   = \big(rep(y_i,m_i),rep(\xb_i,m_i)\big)$ and use $\left(y^M,X^M\right)$ to refer to the 
artificial data set so constricted; that is, $\left[  \beta  | \alpha, M, \Yt, X \right] = \left[  \beta  |  Y=y^M,X^M\right]$\\

Then the vector of observations is $y_i; i= 1,\ldots n$, and each is accompanied by $m_i; i= 1,\ldots n$. This formulation is equivalent, for inference on $\beta$, to the regression of the vector of $y^M$ on $X^M$. Formally $[\beta | \Ybt, \mathbf{M},X] = [\beta | \Yb^{\Mb},X^M]$. Integrating over $\Mb$ leads to 
$$[\beta | \Ybt, \alpha,X] = E_{\Mb | \Ybt, \alpha}\left[  \beta  | \alpha, \Ybt, \Mb ,X^{\mathbf{M}} \right] 
= E_{\Mb | \Ybt, \alpha}[\beta | \Yb^{\mathbf{M}},X^{\mathbf{M}}]$$
Clearly this can be more simply achieved by weighting each case $(y_i, \xb_i)$ by random $M_i$, if that is supported. 
The natural integration method is by Monte Carlo. \\

Another possibility exploits the fact that we can write 
$$[\beta |\Yb^{\Mb}] \propto [\Yb^{\Mb}|\beta] [\beta] = \prod_i \Big[ Y_i^{M_i}| \beta \Big] [\beta] 
= \prod_i [ Y_i, \beta]^{M_i} [\beta]$$
But $ E_{M_i | \IYbt, \alpha} [ Y_i |\beta]^{M_i} = E_{M_i | \Iyi, \alpha} [ Y_i |\beta]^{M_i}$ is analytically available  from the pgf of $[M_i  | \Iyi, \alpha]$. Further this distribution is simple, being concentrated on $M_i =1$ when $y_i >0$; binary when $y_i=0$ and $\omega >0$, and geometric when $y_i=0$ and $\omega < 0$.\\

\subsection{Other distribution families}

There is a considerable interest in disentangling the effects of over-dispersion and zero-inflation. As mentioned, several authors have used distributions such as the negative-binomial, Conway-Poisson-Maxwell and power series distributions for this purpose, for such distributions already have  a parameter for over-inflation. In practice, it seems likely that only in data  sets with a very strong signal will such disentangling be feasible. \\

Two general families are are particularly attractive, but seem not to have been much pursued: the classic exponential-dispersion family, and the exponential over dispersion model of \citet{dey1997overdisp:glm}. These seem to complement the notation of this paper, and specifically, the multiplicative model of ZI. In the former, with $\log(\pi_y(\theta, \eta))= \frac{\theta y - A(\theta)}{\eta}$ to within an additive function of $y$, we may similarly define $\pit_y(\theta, \eta, \gamma) = \omega \Iy + \frac{\theta y - A(\theta)}{\eta}$. In the latter, following  \citet{dey1997overdisp:glm}: $\log(\pi_y(\theta, \zeta)) = \theta y + \zeta Z(y) +\chi(\theta, \zeta)$ is a member of the 2-dimensional exponential family, and thus $\log(\pit_y(\theta, \zeta, \omega)) = \Iy \omega + \theta y + \zeta Z(y) +\chi(\theta, \zeta)$ is a member of the 3-dimensional exponential family if $\omega$ is independent of $(\theta, \zeta)$. Of course, we may go further for this ZI model, for if $\pi_y(\theta)$ for \emph{k}-dimensional $\theta$ has the  form of a \emph{k}-dimensional exponential family distribution, then $\log(\pi_y(\theta, \omega)) =  \omega \Iy + \log(\pi_y(\theta))$ is (\emph{k}+1)-dimensional exponential family.\\
 
\subsection{Multiple and continuous inflations}

We have focussed on zero inflation, that is the inflation of  $\pi_0$ to $\pit_0$.  But some authors \citep[e.g.][]{sweeney2012PhD,deng2015ZANI,tian2015ZANI} in dealing with ZI binomial distributions $Bin(n, p)$, have remarked that the probabilities of both $(y=0)$ and $(y=n)$ can be inflated. Trivially the notation above may be adapted to the inflation of the probability of any other values of $Y$, such as $y=k \in K$ where interest will instead focus on multiple $k$ and specifically on $\pi_k$ and $\pit_k$ for $k \in K$  via 
\begin{equation}\label{multiple}
\log(\pit_y(\theta, \omega)) = \omega_k(K) {\mathbb{I}\{Y = k \in K\}}+\rho + \log(\pi_y(\theta)).
\end{equation}
For example, in the binomial case, with $K = \{0,n\}$,  $ \omega_0( \{0,n\})$ and $\omega_n(\{0,n\})$ control, respectively, the inflations of $\pi_0$ and $\pi_n$. Clearly each can be parameterised via $\gamma_k$ and hence involve covariates.\\

The zero-inflation of a continuous distribution has often  been remarked on \citep[e.g.][]{Lambert1992:ZIP}. The classic example is rainfall; more generally values less than some (possibly small) threshold are returned as zero. If $\pi(y,\theta)$ denotes the pdf of a continuous distribution then with $K$ now denoting a continuous interval
$$\log(\pit (y, \theta, \omega)) = \omega{\mathbb{I}\{y \in K\}} +\rho + \log(\pi(y, \theta))$$
models a single parameter inflation of the pdf $\pi(y, \theta)$ for $y \in K$.  This will have the form of a continuous pdf with a probability mass at $y=0$.\\

However this itself can be extended by $\omega = \omega(y,K)$ taken to be a continuous function. Then the resultant $\pit (y, \theta, \omega)$ will be a conventional continuous pdf, inflated continuously over $y \in K$.

\subsection{Multivariate ZI}

Several authors \citep[e.g.][]{li1999multivarZI,dong2014multivariate,liu2015typeI,lee2017} have reported on the ZI of multivariate counts. The simplest discrete multivariate distribution is the Multinomial. But here the issue is multivariate ZI modelling. For illustrative purposes we consider bivariate counts $y = (y_a, y_b)$, and bivariate $\pi_y(\theta)$. There are three versions of zero: (i) $(0, \cdot), (\cdot, 0), (0,0)$; here for example, $(0, \cdot) \equiv (y_a=0, y_b>0)$. Extending the notation of \eqref{multiple} we have three versions of $\omega:$ being $ \omega_{(0, \cdot)}, \omega_{(\cdot,0)},\omega_{(0, 0)}$. It is often the case that (0,0) is never observed in a multinomial setting.\\

The only new issue is parsimony; for with $k$-dimensional $y$ there are $2^k -1$ such parameters. A simpler version of the above is however available: $\omega_{(0, 0)} = \omega_{(0, \cdot)} + \omega_{(\cdot,0)}$; we have dropped the interaction terms. For higher dimensional cases, we can similarly drop all interactions, or all interactions higher than two-way, etc. The real challenge is the paucity of models $\pi_y(\theta)$ for discrete multivariate data (ie Absent ZI).

\bibliographystyle{agsm}
\bibliography{ZIbibtex}{}

\end{document}